\documentclass[prl,amssymb,twocolumn,superscriptaddress]{revtex4}
\usepackage{graphicx}
\usepackage{amsmath}
\usepackage{bm}

\begin{document}

\title{Non-equilibrium relaxation of an elastic string in a random potential}

\author{Alejandro B. Kolton}
\affiliation{Universit\'e de Gen\`eve, DPMC, 24 Quai Ernest Ansermet,
CH-1211 Gen\`eve 4, Switzerland}
\author{Alberto Rosso}
\affiliation{
Laboratoire de Physique Th\'{e}orique et Mod\`{e}les Statistiques \\
B\^{a}t. 100 Universit\'{e} Paris-Sud;
91405 Orsay Cedex, France}
\author{Thierry Giamarchi}
\affiliation{Universit\'e de Gen\`eve, DPMC, 24 Quai Ernest Ansermet,
CH-1211 Gen\`eve 4, Switzerland}

\begin{abstract}
We study the non--equilibrium motion of an elastic string in a two
dimensional pinning landscape using Langevin dynamics simulations.
The relaxation of a line, initially flat, is characterized by a
growing length, $L(t)$, separating the equilibrated short length
scales from the flat long distance geometry that keep memory of the
initial condition. We show that, in the long time limit, $L(t)$ has
a non--algebraic growth with a universal distribution function. The
distribution function of waiting times is also calculated, and
related to the previous distribution. The barrier distribution is
narrow enough to justify arguments based on scaling of the typical
barrier.
\end{abstract}


\maketitle

The physics of disordered elastic systems has been the focus of
intense activities both on the theoretical and experimental side.
Indeed it is relevant in a large number of experimental situations
ranging from periodic systems such as vortex lattices
\cite{vortex_review_global}, charge density waves
\cite{gruner_revue_cdw}, and Wigner crystals
\cite{giamarchi_wigner_review} to domain walls in magnetic
\cite{lemerle_domainwall_creep,shibauchi_creep_magnetic,caysol_minibridge_domainwall}
or ferroelectric \cite{tybell_ferro_creep,paruch_ferro_exponent}
systems, contact lines \cite{moulinet_contact_line} and fluid
invasion in porous media \cite{wilkinson_invasion}. Because of the
competition between disorder and elasticity, glassy properties
arise, and one of the most challenging question is to understand
their consequences on the dynamics of the system
\cite{giamarchi_sitges_review}.

Since the system must move by thermal activation over the barriers
separating metastable states, the steady state response to a small
external force is a way to probe its glassy nature. The glassiness
leads to divergent barriers and thus to a slow response known as
creep \cite{ioffe_creep,nattermann_rfield_rbond}. Experiments
\cite{lemerle_domainwall_creep,caysol_minibridge_domainwall,tybell_ferro_creep,paruch_ferro_exponent}
as well as microscopic calculations of the response
\cite{chauve_creep,muller_creep_frg} have confirmed this creep
behavior, although questions remain in low dimensions about the
value of the creep exponent \cite{kolton_creep}. Much less is known
about the glassy effects in the case of non-stationary relaxation
towards equilibrium. Understanding such non-stationary physics is
clearly crucial since it gives complementary information on the
barriers and, for experiments, is needed to describe the many
systems that are quenched in the glassy state (e.g. by changing
rapidly the temperature), and have then to relax. Theoretical
attempts to tackle this problem have been made using mean field and
renormalization group approaches
\cite{cugliandolo_relaxmanifold,balents_tbl,schehr_dynamics}. Direct
application of these results to one dimensional domain walls is
however difficult. Numerical studies, that would give more direct
information in low dimension, are also difficult since they have to
deal with ultra long time scales dynamics. Simulations have thus
been mostly restricted so far to 2--dimensional random Ising models
or 2--dimensional periodic elastic systems \cite{paul_puri_rieger,
schehr_2d,rieger_num_2d}. The relaxation of a directed polymer has
been investigated
\cite{yoshino_creep_frg,barrat_vs_yoshino,yoshino_unpublished} by
local Monte Carlo dynamics \cite{rosso_vmc}, but a precise study of
the connection between relaxation and the static glassy properties
is still lacking.
\begin{figure}
 \centerline{\includegraphics[width=8.5cm,height=9.0cm]{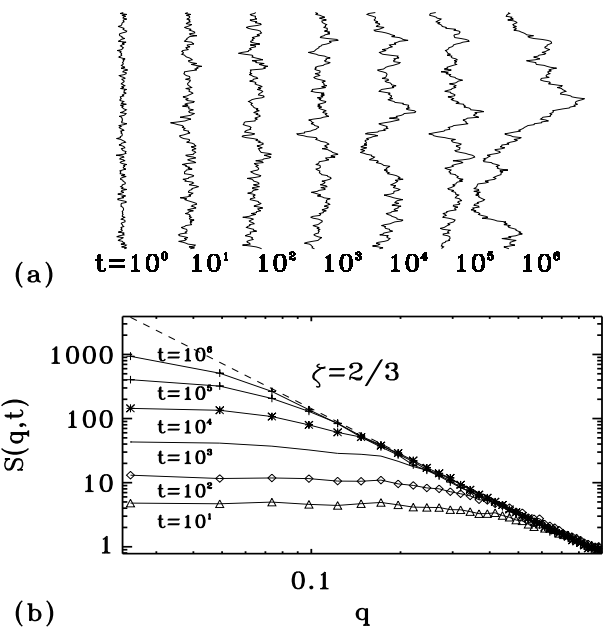}}
 \caption{(a) Typical configurations of the string for
 different times, at $T=0.5$. (b)
 Structure factor for different times
 at $T=0.5$, averaged over $1000$ disorder realizations. The
 dashed line corresponds to the thermal equilibrium
 solution which it is reached at very long times.}
 \label{fig:fig1}
\end{figure}

In this paper we thus study the slow non-equilibrium relaxation of
an elastic string moving in a two dimensional random media. We
prepare the string in a flat configuration and let it relax. We show
that the relaxation is governed by a characteristic growing length,
$L(t)$, separating the equilibrated short length scales from the
flat long distance ones that keep memory of the initial condition.
In the long time limit, $L(t)$ has a non--algebraic growth with a
universal distribution function. We compute the distribution of
waiting times and thus of barriers. This later distribution is found
to be narrow enough to justify the scaling for $L(t)$ based on a
typical barrier.

We consider a string described by a single valued function $u(z,t)$,
measuring its transverse displacement $u$ from the $z$ axis at time
$t$. The initial condition is flat $u(z,t=0)=0$, and we monitor the
relaxation towards equilibrium. The string obeys the equation of
motion:
\begin{equation} \label{eq:motion}
 \gamma \partial_t u_(z,t)= c \partial_z^2 u(z,t) + F_p(u,z) + \eta(z,t)
\end{equation}
where $\gamma$ is the friction coefficient and $c$ the elastic
constant. The pinning force $F_p(u,z)=-\partial_u U(u, z) $ derives
from the random bond disorder potential $U(u, z)$ and the thermal
noise $\eta(z,t)$ satisfies $\langle\eta(z,t)\rangle=0$ and
$\langle\eta(z,t) \eta(z',t')\rangle=2 \gamma T \delta(t-t')
\delta(z-z')$ where $\langle\ldots\rangle$ is the thermal average.
The sample to sample fluctuations of the random potential are given
by $\overline{[U(u,z)-U(u',z')]^2}=-2\delta(z-z') R^2(u-u')$ where
$\overline{\phantom{abc}}$ denotes an average over disorder
realizations. In the random bond case the correlator $R(u)$ is short
ranged.

To solve numerically (\ref{eq:motion}) we use the method of
\cite{kolton_creep}. We discretize the string along the $z$
direction, $z \rightarrow j = 0,\ldots,L-1$, keeping $u_j(t)$ as a
continuous variable. A second order stochastic Runge-Kutta method is
used to integrate the resulting equation. To model a continuous
random potential we generate, for each $j$, a cubic spline
$U(u_j,j)$ passing through regularly spaced uncorrelated Gaussian
random points \cite{rosso_depinning_simulation,kolton_creep}. To
characterize the geometry of the line during the relaxation we
introduce the structure factor $S(q,t) \equiv \overline{s(q,t)} =
\frac{1}{L}\overline{\langle u^*_q u_q \rangle}$ where
$u_q=\sum_{j=0}^{L-1}
 u_j(t) e^{-iqj}$ and $q=2\pi n/L$ with $n=1,\ldots,L-1$.

In the absence of disorder the relaxation of the string can be
solved analytically and $S(q,t)$ is
\begin{equation} \label{eq:Spure}
 S_{pure}(q,t)=S_{pure}^{eq}(q)[1-\exp(-2cq^2 t/\gamma)]
\end{equation}
where $S_{pure}^{eq}(q)= T / c q^2$ is the structure factor at
equilibrium. From (\ref{eq:Spure}) we can separate two regimes: (i)
at large $q$ 
the line is equilibrated with the thermal bath and its geometry is
described by the equilibrium roughness exponent $\zeta=1/2$. This
behavior can be extracted from the $q^{-(1+2\zeta)}$ power law decay
of the structure factor. (ii) at small $q$, however, the string has
still a memory of the flat initial condition, and the structure
factor reaches a plateau: $S_{pure}(q \rightarrow 0^+,t) = (2
T/\gamma) t$. The crossover between these two regimes is driven by a
unique growing characteristic length scale $L(t)$ that can be
defined from the intersection point of the two limiting behaviors.
In the pure case $L_{pure}(t)=2\pi \sqrt{2ct/\gamma}$ and its power
law growth defines the dynamical exponent $z$, as $L(t) \sim
t^{1/z}$.

\begin{figure}
 \centerline{\includegraphics[width=8.5cm,height=9.0cm]{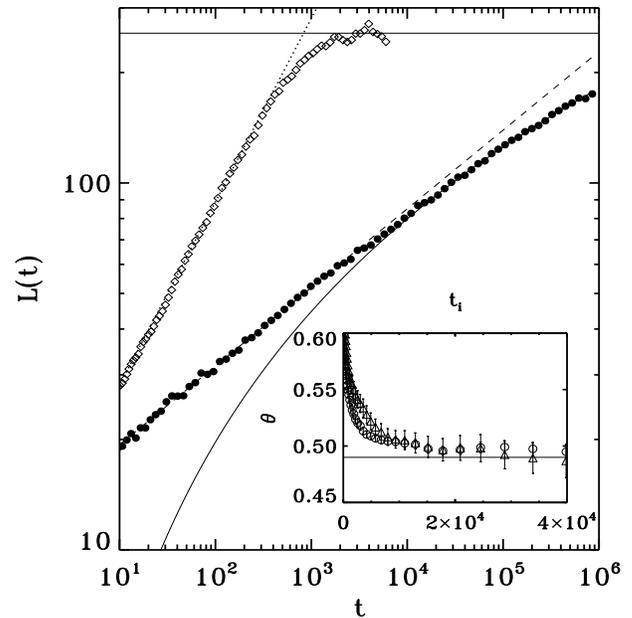}}
 \caption{Growing characteristic length scale $L(t)$ of a string of
 size $L=256$. ($\diamond$) symbols correspond to the relaxation of the
 clean system and the dotted line to the analytical result. ($\bullet$) symbols
 correspond to the disordered case. The continuum line is a
 fit to equation \ref{eq:Lt}, the dashed line is a fit to the power
 law growth at intermediate scales. For $10^4<t<10^6$ we
 get $\theta \approx 0.49$.  Inset: exponent
 $\theta$ extracted from the fit to equation \ref{eq:Lt} in the
 time interval $t_i<t<10^6$. ($\triangle$) symbols correspond to a
 system size $L=512$, and ($\circ$) symbols to $L=256$.}
 \label{fig:fig2}
\end{figure}

We now discuss our numerical results for the system with disorder.
We simulate lines of size $L=256$, $512$, $1024$ with $c=\gamma=1$.
We take $R(0)=1$ and temperatures ranging from $T=0.1$ to $T=0.7$.
In Fig.~\ref{fig:fig1}(a) we show the typical relaxation of a
string. Note that in the pure case, for the same parameters, the
equilibration of a line of size $L=256$ occurs after a time $t \sim
10^3$ (see Fig.~\ref{fig:fig2}). The presence of barriers in the
disordered case makes the dynamics much more slow, and equilibrium
is not yet reached at time $t=10^6$. We show in
Fig.~\ref{fig:fig1}(b) the evolution of $S(q,t)$. As in the pure
case two regimes are observed.  At short length scales the line has
reached equilibrium in the random environment and it is
characterized by the well known roughness exponent $\zeta=2/3$
\cite{kardar_exponent_line}. At large length scales a plateau is
still present and a crossover growing length $L(t)$ can be defined.
Quite generally the scaling form of $S(q,t)$ can be written as,
\begin{equation} \label{eq:scaling_S}
 S(q,t)= S^{eq}(q) G(q L(t))
\end{equation}
where $G(x \rightarrow 0) \sim x^{1+2\zeta}$ and $G(x \rightarrow
\infty) = 1$.
The analytical calculation of $L(t)$ is clearly a non trivial task,
but a simple estimate can be done relying on phenomenological
scaling arguments, based on creep. At low temperatures the
relaxation is dominated by the energy barriers $U(L)$ that must be
overcomed in order to equilibrate the system up to a length scale
$L$. Using the Arrhenius thermal activation law we can thus express
the relaxation time $t(L) \sim \exp[\beta U(L)]$.  Even if the exact
numerical determination of $U(L)$ is an NP-complete problem it is
usually conjectured that the typical barriers of the energy
landscape scale, asymptotically with $L$, the same way as the free
energy fluctuations: $U(L)\sim L^{\theta}$, with $\theta=1/3$ for a
line. Numerical calculations \cite{drossel_barrier} and FRG
calculations \cite{chauve_creep} seem to confirm this conjecture.
Following these arguments we infer that \cite{fisher_droplets}
\begin{equation} \label{eq:Lt}
 L(t) \approx L_c \left[\frac{T}{U_c}\log
 \left(\frac{t}{t_0}\right)\right]^{\frac{1}{\theta}}
\end{equation}
where $L_c$ is a characteristic length which can be identified with
the Larkin length \cite{larkin_ovchinnikov_pinning}, $U_c$ the
associated energy scale $U_c=U(L_c)$ and $t_0$ a microscopic time
scale. An alternative form of $L(t)$ would be the power law scaling
of the clean system, $L(t)\sim t^{1/z}$, but with a new exponent
$z>2$ taking into account the effect of the energy barriers. Note
that this proposal corresponds to thermally activated motion over
barriers scaling logarithmically with the size $L$. Such behavior
has been observed in various 2--dimensional disordered systems
including periodic elastic systems in the so called ``marginal glass
phase''
\cite{schehr_dynamics,paul_puri_rieger,schehr_2d,rieger_num_2d}. For
this model it is possible to show that the dynamical exponent takes
the form $z(T)\propto 1/T$. Moreover the relaxation towards a steady
state of an elastic string just above the depinning threshold shows
the same power law behavior with a dynamical exponent $z<2$
\cite{schehr_dynamics}.

We now compare our results with the above different scenarios. The
growing length scale $L(t)$ can be determined from the average
structure factor $S(q,t)$ shown in Fig.~\ref{fig:fig1}(b). In
practice we define $L(t)$ as the intersection between the plateau
$S_t=S(q \rightarrow 0^+,t)$ and the equilibrated structure factor
$S^{\text{eq}}\sim q^{-7/3}$. The result is shown in
Fig.~\ref{fig:fig2}. Note that the whole time dependence of $L(t)$
is described neither by (\ref{eq:Lt}) neither by a pure power law.
The latter scaling can only approximately fit the short time
relaxation: the fitted dynamical exponent $z$ strongly decreases
with increasing temperature and ranges from $20$ to $4$. However,
for long times, this powerlaw scaling can be ruled out due to the
observed bending in the log-log scale. To be sure that this bending
is not an artifact of the proximity of the finite size equilibration
we verified its presence for bigger systems up to a size $L=1024$
where $L(t)\ll L$ for all considered times. For this reason the
logarithmic growth seems to be more adequate for long times. A
two-parameters fit to (\ref{eq:Lt}) gives an exponent $\theta$
which, at long times, becomes size and time independent, as shown in
the inset of Fig.~\ref{fig:fig2}.

Although the logarithmic growth law describes well our data at long
times, we find an exponent $\theta \approx 0.49$, bigger than the
expected value $1/3$. If we assume that the dynamics of relaxation
is governed by Arrhenius activation, this result indicates either a
violation of the expected scaling of barriers or the presence of
non-negligible sub-leading corrections in this scaling at the length
scales spanned by $L(t)$ in our simulations. The inset of
Fig.~\ref{fig:fig2} shows, for different time-windows, the exponent
$\theta$. The saturation of $\theta$ excludes strong sub-leading
corrections at least for the largest times reached in the
simulations. However, the adequacy of the fit with $\theta \approx
0.49$ in the last three decades is still not enough to exclude the
presence of logarithmic corrections in the leading term: $U(L)\sim
L^{1/3} \log^{\mu}(L)$. The latter scenario is consistent with the
upper bound scaling found numerically in \cite{drossel_barrier} for
the barriers separating metastable states of a directed polymer in
2--dimensional random media. Such a scaling has been shown
\cite{yoshino_unpublished} to also fit well the Monte Carlo
relaxation data for a directed polymer.

The scaling of the barriers $U(L)$ and the subsequent evolution of
$L(t)$ refer to {\it typical} values of $U$ and $L(t)$. On the other
hand, for broad enough distributions typical and mean values can be
very different \cite{vinokur_marchetti}. Therefore, the deviations
of the numerical data from the predicted behavior (\ref{eq:Lt})
might be produced by a broad distribution of barriers. To check for
such a possibility, and to extract the barrier distribution, we
study the sample to sample fluctuations of the various observables.
A convenient quantity to compute for each evolving sample is the
instantaneous value of the structure factor plateau $s=s(q
\rightarrow 0^+,t)$ wich is directly related to the growing length
$l \sim s^{1/(1+2\zeta)}$. As raw data directly confirms, this
quantity is a stochastic process growing monotonically with the time
$t$. Thus,
 its sample to sample fluctuations can be directly related to the
distribution of relaxation times $\tau$ and to the statistics of
barriers $u$ by assuming Arrhenius activation, $u \sim \log(\tau)$
\footnote{Strictly speaking $u$ is the logarithm of the {\it total}
waiting time $\tau$ to equilibrate a fixed length scale $l$.
However, if as usually advocated, $\tau$ is dominated by the slowest
Arrhenius activated processes, then the quantity $u$ we obtain is
indeed the barrier at the lengthscale $l$, $\tau \sim e^{\beta
u(l)}$ \cite{kolton_relaxation_long}. We thus use this denomination,
even if it is slightly improper, in what follows.}. One obtains
\cite{kolton_relaxation_long}
\begin{equation}\label{eq:continuity}
 \Phi_s(u)=1-\Phi_u(s)
\end{equation}
where $u$ (resp. $s$) is the sample dependent barrier (resp.
structure factor plateau) and $\Phi_s(u)$ ($\Phi_u(s)$) its
cumulative distribution function for a given value of $s$ (resp.
$u$) (i.e., $\Phi_s(u_0)$ is the probability to find a barrier $u$
smaller than $u_0$, given a fixed value $s$ for the plateau).
\begin{figure}
 \centerline{\includegraphics[width=8.5cm,height=9.0cm]{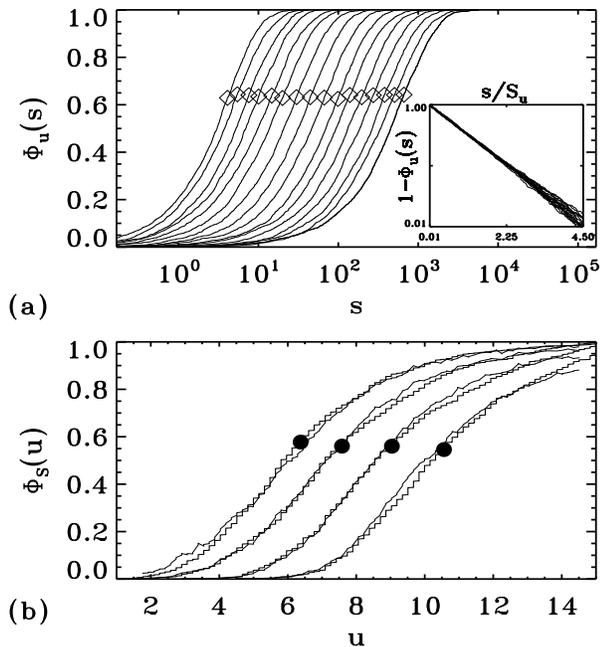}}
 \caption{(a) Cumulative distribution function of the plateau
 value $s$ for fixed times $t$ ranging
 from $5$ to $10^6$. ($\diamond$) symbols indicate the mean
 value $S_t$. Inset: collapse of the cumulative
 distributions in the rescaled variable
 $s/S_t$. (b) Cumulative distribution function of
 the barriers $u$ for $s=10, 20, 40, 80$.
 Circles ($\bullet$) indicate the mean value $U_s$. Step-lines are
 obtained from raw data, while solid lines are obtained from
 the rescaled cumulative distribution of $s$, (\ref{eq:continuity}).
 }
 \label{fig:fig3}
\end{figure}
Fig.~\ref{fig:fig3}(a) shows $\Phi_u(s)$ as a function of $s$ for
different values of $u$. For all $u$ the distributions are narrow
and, on a logarithmic scale, appear just shifted. This suggest the
simple rescaling $s/S_t$, which collapses all the curves as shown in
the inset. Strikingly, we find that this rescaled function for the
fully disordered system is indistinguishable, at the resolution of
our numerical study, from the one ($\Phi_u(x=s/S_t)= 1-\exp(-x)$) of
the clean system, and from the identical one would obtain for the
Larkin model \cite{larkin_70} of disorder (despite the fact that
this model does not have pinning and metastable states). This
scaling form implies that the sample to sample fluctuations of the
growing length, $l(t)$, are given by $\Phi_u(x=l/L_t)=
1-\exp[-\alpha x^{1+2\zeta}]$, with
$\alpha=\Gamma\bigl(1+\frac{1}{1+2\zeta}\bigr)^{1+2\zeta}$. The
statistics of barriers is obtained from (\ref{eq:continuity}) using
the evolution $S_t$ vs $u \sim \log(t)$ of Fig.~\ref{fig:fig2}. In
Fig.~\ref{fig:fig3}(b) we show that the cumulative distribution
$\Phi_u(s)$ derived using the latter method indeed coincides with
the one obtained from a direct analysis of the raw data of $s$ vs
$u$ for each sample. As for the sample dependent plateau $s$ (for
given values of $u$), the distributions of $u$ for given values of
$s$ are found to be exponentially narrow. Scaling arguments based on
typical values are therefore justified, since they can be safely
translated directly to the mean values. This indicates that the
effect of sample to sample fluctuations cannot explain the
deviations of the numerical data with respect to the
phenomenological predictions observed in Fig.~\ref{fig:fig2}, and
that such deviations must come from the scaling of the barriers.
Note also that, as visible in Fig.~\ref{fig:fig3}(b), the barrier
distribution $\Phi_s(u)$, contrarily to the distribution of plateaux
$\Phi_u(s)$, does {\it not} scale with $u/U_s$, where $U_s =
\overline{\langle u \rangle}$ is the mean value. Such a scaling
would only work if a {\it pure} power law scaling of the barriers
with length were perfectly verified. The complex behavior of
$\Phi_s(u)$ clearly comes from the existence of two regimes in the
scaling of the barriers as a function of time (length) as shown in
Fig.~\ref{fig:fig2}. However, it remains to be understood why the
presence of these two regimes does not affect the perfect collapse
for $\Phi_u(s)$ as a function of $s/S_t$, ranging from the shortest 
to the longest times. No analytical explanation of this fact, nor of
the form of the corresponding scaling function exists so far.

At long time, approximate power-law scaling for the barriers is
recovered, and thus the distribution of barriers would scale with
$u/U_s$. In this case (\ref{eq:continuity}) shows that the universal
function for all the cumulative distribution functions $\Phi_s(u)$
would be a stretched exponential. This form is different from the
one that extremal statistics arguments would suggest
\cite{vinokur_marchetti}, prompting for a reexamination of the
physical understanding of the barrier distribution in such
disordered systems.

We acknowledge discussions with L. Cugliandolo, P. Le Doussal, and G.
Schehr. We acknowledge H. Yoshino for discussions and 
for making Ref. \onlinecite{yoshino_unpublished} available before 
publication. This work was supported in part by the Swiss
National Fund under Division II.


\end{document}